\documentstyle[11pt,moriond,epsfig]{article}

\def\be{\begin{equation}}
\def\ee{\end{equation}}
\def\bea{\begin{eqnarray}}
\def\eea{\end{eqnarray}}

\newcommand{\beastar}{\begin{eqnarray*}\extraspace} 
\newcommand{\eeastar}{\end{eqnarray*}}

\newcommand{\eps}{\epsilon}

\newcommand{\bra}{\langle}
\newcommand{\ket}{\rangle}

\begin{document}
\title{Quasi-particles for quantum Hall edges \footnote{ Contribution to the proceedings of the XXXIVth
  Rencontres de Moriond `Quantum Physics at Mesoscopic Scale'}}
\author{ K. Schoutens and R.A.J. van Elburg }

\address{Institute for Theoretical Physics and
               Van der Waals-Zeeman Institute\\
              Valckenierstraat 65, 1018 XE Amsterdam,
              The Netherlands} 

\maketitle\abstracts{We discuss a quasi-particle formulation of 
effective edge  theories for the fractional quantum Hall effect.  
Fundamental quasi-particles for the Laughlin state with filling 
fraction $\nu={1 \over 3}$ are edge electrons of charge $-e$ 
and edge quasi-holes of charge $+{e \over 3}$. These quasi-particles 
satisfy exclusion statistics in the sense of Haldane. We exploit 
algebraic properties of edge electrons to derive a kinetic equation 
for charge transport between a $\nu={1 \over 3}$ fractional 
quantum Hall edge and a normal metal.}

\section{Introduction}

In the first paper \cite{TSG} on the fractional quantum Hall effect,
Tsui, St\"ormer, and Gossard already suggested the existence of
fractionally charged quasi-particles over fractional quantum Hall
states. They extended Laughlin's gauge argument from the integer to
the fractional quantum Hall effect (fqHe) and argued that fractionally
charged quasi-particles could be expected. Soon after that Laughlin
constructed an approximate ground state wave function and argued using
Schrieffer's counting argument that a fractionally charged
quasi-particle could be constructed by piercing the ground state with
an infinitely thin solenoid and adiabatically inserting a flux quantum
through this solenoid \cite{La}. 

In view of the above, the existence of the fqHe at simple filling 
fractions is equivalent to the existence
of fractional charge. In the last few years more direct measurements
of the quasi-particle charge have been performed.  Goldman et
al. \cite{G2} used a quantum antidot as an electrometer to measure
the charge of the excitations in the $\nu = {1\over 3}$ fractional  
quantum Hall state.  At this conference L.~Saminadayar discussed recent
shot-noise experiments performed by Saminadayar et al. \cite{SGJE}
and de Picciotto et al. \cite{PRHUBM}, which showed that the
tunneling current from one $\nu={1\over 3}$ quantum Hall edge to an
other is carried by quasi-particles of fractional charge ${e\over 3}$ .  
M.~Reznikov reported the observation of charge ${e\over 5}$
quasi-particles in a similar shot-noise experiment on a $\nu={2\over 5}$
fractional quantum Hall system. 

In the light of these fascinating observations of fractional charge, it 
might come as a surprise that in most 
theoretical work on the quantum Hall effect bosonization schemes are
used, in which the low energy edge excitations of the fractional
quantum Hall system are described by neutral bosonic excitations. At
this conference, we reported on an alternative approach 
that gives a central role to (fractionally) charged quasi-particles at 
the 
edge of a fractional quantum Hall system, and we discussed the fractional 
exclusion statistics of these quasi-particles. In a most interesting 
contribution, S.B.~Isakov proposed how the fractional statistics of 
quantum Hall quasi-particles can be used for the analysis of shot-noise 
experiments. 

In the now following sections we briefly discuss the properties of charged 
quasi-particles at a $\nu={1\over3}$ fractional quantum Hall edge, with 
special emphasis on the statistics properties. We refer to our paper 
\cite{vES} for a more detailed discussion. 

\newpage

\section{Charged quasi-particles at a  $\nu={1 \over 3}$ edge.}

\subsection{Hall conductance}

Before giving any further details, we present a quick argument that 
illustrates the necessity of assigning fractional exclusion statistics 
to charged quasi-particles in fqHe edges. 
One quickly checks that, in an effective edge description, the zero 
temperature Hall conductance $\sigma_H$ is expressed as
\be 
\sigma_H = n^{\rm max} \, {q^2 \over h} 
\label{sigmaH}
\ee
with $q$ the charge of the quasi-particles that carry the edge current and 
$n^{\rm max}$ the maximum value of the thermodynamic distribution 
function of these same quasi-particles. For the $\nu={1 \over 3}$ edge,
with $\sigma_H={1 \over 3}{e^2 \over h}$, both the charge $q=-e$ and
charge $q={e \over 3}$ quasi-particles are seen to have 
$n^{\rm max} \neq 1$, implying that both types of quasi-particles are 
described by exclusion statistics different from Fermi statistics.

\subsection{Edge electron states}

Before turning our attention to quasi-particles with fractional charge,
we first discuss quasi-particles with the charge of an electron and 
fermion-like exchange statistics. We call these quasi-particles edge 
electrons.

The starting point for this analysis are operators
$\Psi^{\pm}(z)$ that describe the creation and annihilation of edge
electrons in a second quantized field theory \cite{We1}.  
This field
theory is a so-called Conformal Field Theory, and in what follows we
shall exploit special algebraic properties associated to conformal
invariance. We use the mode expansion
$\Psi^{\pm}(z)=\sum_t \Psi^{\pm}_{t} z^{-t-{3 \over 2}}$. The mode
index $t$ takes half-integer values and corresponds to the
dimensionless energy of the mode, i.e., $\eps_t=t {2\pi \over
L}{1\over \rho_0}$ with $\rho_0$ the density of states per unit
length, $\rho_0=(\hbar v_F)^{-1}$ and $L$ the length of the
edge. Calling $\Psi^\dagger_t=\Psi^-_{-t}$ and $\Psi_t=\Psi^+_t$, we
identify $\Psi^\dagger_t$ and $\Psi_t$, with $t>0$, with the creation
and annihilation operators of an edge-electron of energy $\eps_t$,
respectively.  These operators satisfy the anti-commutation rules
\be
\{ \Psi^\dagger_r, \Psi_s\}= (r^2-{1\over 4})\delta_{-r+s}+ 6\, 
L_{-r+s} +3(r+s) J_{-r+s}.
\label{commrel}
\ee
In this relation, $L_{m}$ is the $m$-th Fourier mode of the 
energy momentum tensor and similarly $J_m$ is a Fourier mode of the 
current. The zero modes have a simple meaning: $L_0$ is the hamiltonian 
and $J_0$ measures the charge. The important point to notice here is 
that for a $\nu=1$ integer quantum Hall edge, which is a Fermi-liquid, the
anti-commutation relations
\be
\{ \Psi^\dagger_{r,\nu=1}, \Psi_{s,\nu=1}\}= \delta_{-r+s}
\label{commrel1}
\ee
are much simpler and only involve the edge electron operators
themselves. In a fractional quantum Hall edge, however, the
anti-commutator is a non-trivial operator whose value depends on
the state in which the expression is evaluated.

By repeatedly acting with the creation operators $\Psi^\dagger_t$, one 
produces states with more than one edge electron. We will 
now use the anti-commutation relations to show how
some of these states have zero norm.  
The norm-squared of the one-particle state created by
$\Psi^\dagger_t$  is given by
\bea
\langle 0|\Psi_t \Psi^\dagger_t |0 \rangle = 
\langle 0 | (t^2-1/4)+ 6 \, L_0 + 6t \,  J_0 |0\rangle.
\eea
The zero modes $L_0$, $J_0$ evaluated on the vacuum give zero, 
and we see that $t={1\over 2}$ leads to a state with norm zero, and
that the lowest energy one edge-electron state is
$\Psi^\dagger_{3\over 2} 
|0\rangle$. Continuing,
one may try to add  a second quasi-particle to this state.  
A calculation using the full 
algebra satisfied by $\Psi^{\dagger}_r$, $\Psi_s$, $L_m$ and $J_n$
(which is a so-called $N=2$ superconformal algebra)
shows that the  states
\be 
\Psi^\dagger_{1\over 2}\Psi^\dagger_{3\over 2} |0\rangle \, \;\; \;
\Psi^\dagger_{3\over 2}\Psi^\dagger_{3\over 2} |0\rangle  \, \; \;\;
\Psi^\dagger_{5\over 2}\Psi^\dagger_{3\over 2} |0\rangle  \, \;\;\;
\Psi^\dagger_{7\over 2}\Psi^\dagger_{3\over 2} |0\rangle
\ee
have zero norm. The lowest-energy two-particle state with non-zero norm is 
$ \Psi^\dagger_{9\over 2}\Psi^\dagger_{3\over 2} |0\rangle .$\\
Here one sees the start of a pattern: upon adding a third 
quasi-particle, the first two energy levels directly above $t={9\over 2}$ 
will 
be inaccessible, etc. The lowest-energy state  with a total of $M$ 
edge electrons will employ the 1-particle energies $t={3 \over 2},
{9\over 2},\ldots, (2M-1)\, {3 \over 2}$. This means that, of all
allowed 1-particle states, at the most one out of three can be filled.

A complete basis of multi-electron states is given by
\be
 \Psi_{(2M-1){3 \over 2}+ m_M}^\dagger \ldots
 \Psi_{{9 \over 2}+ m_2} ^\dagger
 \Psi_{{3 \over 2} +m_1} ^\dagger| \, 0 \, \rangle
 \qquad {\rm with}\ \quad 
 m_M \geq  \ldots \geq m_2 \geq m_1 \geq 0 \ .
\ee
To this collection of states one may associate a partition function
$Z(\mu_k^e,\beta)$, with $\mu_k^e$ a chemical potential for the $k$-th
one-particle level. Using a method based on the analysis of truncated 
partition sums \cite{Sc,vES}, one shows that in the thermodynamic limit the
partition sum $Z(\mu_k^e,\beta)$ factorizes as a product 
\be
Z(\mu_k^e,\beta)=\prod_k \Lambda_k \qquad {\rm with}\quad 
(\Lambda_k - 1)\Lambda_k^2 = \exp(-\beta(\eps_k-\mu_k^e)) \  .
\label{iow-e}
\ee
Clearly, the quantity $\Lambda_k$ can be viewed as the single-level 
partition sum associated to the $k$-th one-particle energy level.
For non-interacting fermions, the analogous factorization is 
\be
Z_F(\mu_k,\beta)=\prod_k [1+ \exp(-\beta(\eps_k-\mu_k^e))]
\ee
and we recognize the single-level partition sum
$[1+\exp(-\beta(\eps_k-\mu_k^e))]$
 as the one dictated by the Pauli
principle.  The expected occupation of the $k$-th level is given
by the Fermi-Dirac distribution
\be
n_{FD}(k)={1\over \beta}{\partial_{\mu_k^e} Z_F\over Z_F}=
                            {1 \over 1+\exp(\beta(\eps_k-\mu_k^e))} \ .
\ee
In a similar manner, we can read off from the $\Lambda_k$ the
distribution of the $\nu={1\over 3}$ edge-electrons
\be
  n_e(k) \equiv {1\over \beta} {\partial_{\mu_k^e} \Lambda_k \over 
  \Lambda_k} \ .
\label{iow}
\ee
Instead of trying to solve the characteristic equations (\ref{iow-e}),  
we can derive from them 
\be
  n_e(k) = {1 \over 3+ w_k} \ ,
  \quad 
  w_k = (\Lambda_k-1)^{-1} \ .  
\label{dis-e}
\ee
The equations (\ref{iow-e}), (\ref{dis-e}) agree with the Isakov-Ouvry-Wu 
\cite{IOW} equations describing the thermodynamics associated to fractional 
exclusion statistics as defined by Haldane \cite{Ha1}, with statistics 
parameter equal to $g=3$.

\subsection{Quasi-hole states}

By using a similar reasoning, one can analyze the quasi-hole operator
$\phi(z)$ associated with the creation of a fractional charge $q=+{e
  \over 3}$ at the $\nu={1 \over 3}$ edge.
A basis for the corresponding quasi-hole states is 
\be
 \phi_{-{(2N-1) \over 6}- n_N} \ldots
 \phi_{-{3 \over 6}- n_2} 
 \phi_{-{1 \over 6}- n_1} | \, 0 \, \rangle
  \qquad {\rm with}\ \quad 
  n_N  \geq \ldots \geq n_2 \geq n_1 \geq 0 \ . 
\label{phistates}
\ee
The partition sum for these quasi-hole states is again factorizable in the
thermodynamic limit, 
\be
Z(\mu_l^\phi,\beta)=\prod_l \lambda_l \qquad {\rm with}\quad
(\lambda_l -1)^3  \lambda_l^2 = \exp(-3\beta(\eps_l-\mu_l^\phi))
\ee
and we obtain the distribution function
\be
  n_{\phi}(l) = {1 \over {1 \over 3} + w_l},
  \quad 
  w_l = (\lambda_l-1)^{-1} \ .  
\label{dis-phi}
\ee
These relations are equivalent to an IOW equation for Haldane statistics,
this time with the statistics parameter taking the value $g={1\over 3}$.

\subsection{Duality}

Having explained the appearance of the distribution functions for 
fractional exclusion statistics with $g=3$ and $g={1 \over 3}$, 
respectively, we recall that there is a particle-hole duality between 
the two cases \cite{NW,Ra,vES}.
\be
3 \, n_e(\eps) = 1 - {1\over 3} \, n_{\phi}(-{1\over 3} \eps) \ .
\label{dual}
\ee
In our paper \cite{vES}, we demonstrated how a complete basis for the
$\nu={1\over 3}$ edge theory can be obtained by {\it independently}
filling the one quasi-particle spectra of the edge-electron and the 
quasi-hole.
The interpretation of the duality relation is now that the 
positive-energy quasi-hole excitations can be viewed as holes in 
the ground state distribution of
negative energy edge-electrons and vice versa. The relative 
factor $(-{1\over3})$ between the energy arguments in (\ref{dual}) 
indicates that the act of taking out a single edge-electron from a 
filled sea 
corresponds to creating three quasi-holes. 

\section{Transport properties}

The thermodynamic distribution functions that we described satisfy 
$n^{\rm max}_e = {1 \over 3}$ and $n^{\rm max}_{\phi} = 3$, in agreement 
with the expression  (\ref{sigmaH}) for the Hall conductance. Other 
thermodynamic quantities, including the edge contribution to the specific 
heat, are quickly computed in the quasi-
particle formalism. Using Rajagopal's formula for fluctuations \cite{Ra}, 
we also reproduced the expressions for the Johnson-Nyquist noise \cite{Un}. 
Here we do not discuss these results, but move on and consider transport 
properties. 

Following the set-up of a number of recent experiments, we
consider a situation where electrons (or holes) from a
Fermi-liquid reservoir are allowed to tunnel into a $\nu={1 \over 3}$
fqHe edge. The DC $I$-$V$ characteristics for this set-up, which were
first computed by Kane and Fisher \cite{KF} (see also \cite{We2}),
show a cross-over from a linear (thermal) regime into a power-law
behavior at high voltages and thus presents a clear fingerprint of the
Luttinger liquid features of the fqHe edge. The experimental results
from \cite{CPW} are in agreement with these predictions. (We refer 
to the contributions of A.M.~Chang and E.~Fradkin to this conference 
and to \cite{dCF} for a further theoretical analysis.)
The calculations were based on bosonization and on
the Keldysh formalism for non-equilibrium transport. Here we reproduce 
these results in an approach directly based
on the edge quasi-particle formalism.

A careful derivation, based directly on the form of the tunneling
hamiltonian
\be 
H_{int} \propto t \, \int d\eps \, \left[\Psi_{\nu=1}^{\dagger}(\eps) 
\Psi_{\nu={1\over 3}}(\epsilon) + {\rm h.c.} \right] \ , 
\ee 
leads to the following kinetic equation (compare with \cite{We2}) 
\be 
I(V,T) \propto e \, t^2 \int_{-\infty}^\infty d\epsilon 
\left[ f(\epsilon-eV)H(\epsilon)-F(\epsilon-eV)h(\epsilon)\right] \ ,
\label{current}
\ee
where $h,H$ are one particle Green's functions 
\bea
h(\epsilon)=\langle\Psi_{\nu={1\over 3}}^\dagger(\eps)
\Psi_{\nu={1\over 3}}(\eps) \rangle_{V,T}\; , \;\;\; 
H(\epsilon)=\langle \Psi_{\nu={1\over 3}}(\eps)
\Psi_{\nu={1\over 3}}^\dagger(\eps) \rangle_{V,T}
\label{green1}
\eea
for edge electrons in the $\nu={1\over 3}$ fqHe edge, taken at
$V=0$. Here  $f(\eps)$ and $F(\eps)$ are the Fermi-Dirac distributions for 
electrons and holes, respectively. 
The quantities $H(\eps)$ and $h(\eps)$ can be determined as follows.
The ratio of $H(\eps)$ and $h(\eps)$ is fixed,
\be
H(\epsilon)=e^{\beta(\epsilon-eV)}h(\epsilon) \ ,
\ee  
 by {\em detailed balance},
which can be phrased as the requirement that at zero voltage there 
should 
be no current flowing.
The sum $H(\eps)+h(\eps)$ is fixed by the  anti-commutation
relation (\ref{commrel}), here in the continuum approximation
\be
\left\{  \Psi_{\nu={1\over 3}}^\dagger(\eps)
,\Psi_{\nu={1\over 3}}(\eps^{\prime}) \right\} 
= {2\pi \over L} {1 \over \rho_0} \epsilon^2 \delta(\eps-
\eps^{\prime}) 
  + 6 {E_{\eps^{\prime}-\eps} \over \rho_0} + 3 (\eps+\eps^{\prime}) 
{ Q_{\eps^{\prime}-\eps} \over e 
\rho_0} \ .
\label{N2alg}
\ee
In this formula, $E_0$ is the operator for the total energy per unit 
length (proportional to $L_0$), and $ Q_0$ is the operator for the total 
charge per unit length (proportional to $J_0$).
The expectation values of energy and charge can now be calculated using 
the
distribution function $n_e(\eps)$, given in (\ref{dis-e}). We find
\be
\bra E_0 \ket_{V,T} = \rho_0 \left( {\pi^2 \over 6 \beta^2} + 
{(eV)^2 \over 6} \right) \ , \qquad
\bra  Q_0 \ket_{V,T} = - e \rho_0 {(eV) \over 3} 
\ee
and obtain the exact expressions
\be
H(\epsilon)={ (\epsilon-eV)^2+{\pi^2\over \beta^2} \over 
                       e^{-\beta(\epsilon-eV)}+1 }
\ , \quad
h(\epsilon)={ (\epsilon-eV)^2+{\pi^2\over \beta^2} \over 
                      1+e^{\beta(\epsilon-eV)} } \  .
\ee
They lead to $I$-$V$ characteristics
\be 
I(V,T)\propto e \, t^2 \, \beta^{-3} \left({\beta eV\over 2\pi}+
\left({\beta eV\over 2\pi}\right)^3\right) \ ,
\ee
in agreement with the result obtained in different approaches 
\cite{KF,We2}. 

\section*{Acknowledgments}
We thank the organizers of the XXXIVth Rencontres de Moriond `Quantum
Physics at Mesoscopic Scale' for putting together a most
interesting program. This research was
supported in part by the foundation FOM.

\end{document}